\documentstyle[PASJadd]{PASJ95}
\begin{document}
\setcounter{page}{000}

\title{Period Measurement of AGB Stars \\in the Outer Galactic Disk}

\author{Jun-ichi {\sc Nakashima},$^{1,2}$
B.W. {\sc Jiang},$^{1,3}$ Shuji {\sc Deguchi},$^4$ \\
Kozo {\sc Sadakane},$^2$ and Yoshikazu {\sc Nakada}$^5$
\\[12pt]
$^1$ {\it Department of Astronomical Science, The Graduate University for 
Advanced Studies,}\\
{\it Nobeyama Radio Observatory, Minamimaki, Minamisaku, Nagano 384-1305} \\
{\it E-mail(JN): junichi@nro.nao.ac.jp}\\  
$^2$ {\it Astronomical Institute, Osaka Kyoiku University,}\\ 
{\it Asahigaoka, Kashiwara, Osaka 582-8582}\\
$^3$ {\it Beijing Astronomical Observatory, Chinese Academy of Science,}\\
{\it Beijing 100012,P.R.China}\\
$^4$ {\it Nobeyama Radio Observatory, Minamimaki, Minamisaku, Nagano 384-1305}\\
$^5$ {\it Kiso Observatory, Institute of Astronomy, Faculty of Science,}\\
{\it the University of Tokyo, Mitaka, Tokyo 181-8588}}

\abst{Light variation of the 47 AGB star candidates in the outer Galactic disk has 
been monitored at I--band for 5 years. Periods were determined well for 18 of them 
and less reliably for the other 25. The average period of the objects is then 500 
days. According to the period--luminosity relation, the mean luminosity of the sample 
stars is 10000 $\LO$. Based on the absolute luminosity derived from the 
period--luminosity relation and the apparent luminosity from the observation, the 
distances to the objects are determined. The distances calculated are slightly greater 
than those obtained previously on the assumption of constant luminosity of 8000 $\LO$. 
In addition, SiO maser emission was detected for most of the observed objects so that 
their radial velocities were known accurately. With the assumption of circular rotation 
in the Galactic disk, the rotation curve and Oort's constants were derived.}

\kword{Stars:distance --- Stars:late-type --- Stars:
long-period variables --- Stars:stellar dynamics 
--- Galaxy:kinematics and dynamics}

\maketitle
\thispagestyle{headings}

\section
{Introduction}

Up to now, kinematical investigations of disk population stars have been made for 
groups such as OB stars (Fich et al. 1989; Brand \& Blitz 1993), cepheids 
(Pont et al. 1994), carbon stars (Metzger \& Schchter 1994), young open clusters 
(Hron 1987), planetary nebulas (Schneider \& Terzian 1983). However, because of 
interstellar extinction in the Galactic disk at optical and near-infrared wavelengths, 
such work has been limited to regions near the Sun.

In contrast to the above nearby disk-stars, a sample of color-selected IRAS sources, 
the Asymptotic Giant Branch (AGB) star candidates, reaches a distance of about 15 
$kpc$ from the Sun. These color-selected IRAS AGB candidates very often exhibit 
OH/SiO maser emission (te Lintel-Hekkert et al. 1989; Izumiura et al. 1994; Jiang 
et al. 1997); accurate radial velocities of stars have been known from the OH/SiO 
maser observation. Systematic SiO maser searches for AGB star candidates have been 
made in a wide galactic-longitude range, and a large number of radial velocities 
have been accumulated (Izumiura et al. 1994; Jiang et al. 1997). If the distances 
of the AGB candidates are known, the radial velocity data can be used to obtain the 
circular velocity of the Galaxy. In our previous work, distances to the IRAS AGB 
candidates were estimated on the assumption of a constant luminosity, 8000 $\LO$ 
(Jiang et al. 1997; Deguchi et al. 1998). However, once its period of light variation 
is known, the luminosity of an individual star can be determined from the 
period--luminosity relation of Mira-type variables (Feast et al. 1989; Hughes \& 
Wood 1990). Using period information, we thus expect on improvement in the accuracy 
with which the fundamental constants of Galactic kinematics can be found.

In this paper, we report a result of the 5-year monitoring observations of light 
variation in the I--band for 47 objects. In the I--band, AGB variables have large 
variation amplitude compared with those at J, H, K. The period can be found relatively 
easily. We obtained well-determined pulsation periods for 18 objects. Distances were 
then estimated from the period--luminosity relation by the two different methods (see 
section 3.2). Estimated distances are compared with those obtained in the previous 
work. The observed sources are located at $l = 90^{\circ}$ to $230^{\circ}$, i.e., 
in the outer disk region of the Galaxy, where the presence of halo dark matter is 
suspected from the flat rotation curve. We plot the rotation curve of the outer disk 
using our sample. In the past, some the period measurements were done, in the direction 
of Bulge (Glass et al. 1995; Whitelock et al. 1991), and the south polar cap (Whitelock 
et al. 1994). The observations in this paper are the first systematic period measurement 
of Miras on the direction of the outer disk of the Galaxy.

\section
{Observation and Data Reduction}

The photometric monitoring observations were carried out using the 105/150 cm Schmidt 
telescope at Kiso Observatory, University of Tokyo during 1994 - 1997, and the 51 cm 
Cassegrain telescope at Osaka Kyoiku University during 1997 - 1998. The CCD camera 
which was attached to the Schmidt telescope at Kiso Observatory contained a TI Japan 
TC215 chip with an array size 1024 $\times$ 1024 pixsels. The field of view was 
$12.\hspace{-2pt}'5$ $\times$ $12.\hspace{-2pt}'5$ and one pixsel was $0.\hspace{-2pt}''75$ 
on the sky. The CCD images were taken in both V and I filters for every source at Kiso 
Observatory. The limiting magnitudes at Kiso were about 20 mag in the V--band and 19 mag 
in the I--band. In the Kiso photometric system, the V--filter was the same as that in the 
Johnson system, but the I--filter was centered at 8200~\AA~, bluer than 9000~\AA~ of the 
Johnson system (Jiang 1997). The telescope at Osaka Kyoiku University had F/12 Cassegrain 
focus and the detector was a liquid--nitrogen cooled CCD camera using an EEV88200 chip 
with an array size of 1152 $\times$ 770 pixels. The camera had a field of view 
of $14.\hspace{-2pt}'8$ $\times$ $10.\hspace{-2pt}'1$, and it was equipped with Johnson V and 
Cousin R and I interference filters. We used the 2 $\times$ 2 binning mode 
($1.\hspace{-2pt}''54$ pixsels on the sky) throughout the observation at Osaka Kyoiku 
University. The limiting magnitude at Osaka Kyoiku University was about 18 mag in the 
I--band. The CCD images were taken only in one band (I--band) at Osaka Kyoiku University 
except for calibration images which were taken in both R and I filters. 

The observations were done at Kiso from July 1994 to January 1997, and at Osaka from May 
1997 to November 1998. During the observations at Osaka, we made a color-calibration 
observation at both Kiso (at September 1997) and Osaka (from September 1998 to November 
1998). The observation log is given in the Table 1 where the IRAS name and Julian dates 
of observations for all the sources are listed. The seeing sizes were typically 
$3\hspace{-2pt}''$ at both Kiso and Osaka. 

For the purpose of studying the stellar kinematics and the Galactic rotation, individual 
light variation periods (to determine the absolute luminosities using period--luminosity 
relation) and radial velocities are required. Therefore, we needed to choose the candidate  
Mira-type variables and@SiO maser emitters according to suitable criteria. The sources  
in the present paper are taken from the list of IRAS sources selected by Jiang et al.
(1996). They are the AGB star candidates and are identified in the I--band. They were 
chosen in terms of 1) position (Galactic longitude@between $90^{\circ}$ and $230^{\circ}$, 
Galactic latitude between $-10^{\circ}$ and $10^{\circ}$), 2) the flux qualities (333 at 
12, 25 and 60 $\mu$m), 3) the IRAS 12 $\mu$m flux density (brighter than 3 Jy), 4) the 
color $C_{12}~\equiv~log_{10}\left(F_{25}/F_{12}\right)$, where $F_{12}$ and $F_{25}$ 
are the IRAS flux density at 12 $\mu$m and 25 $\mu$m, (between $-$0.3 and 0.3), 5) the 
IRAS variability index (larger than 50, Beichman et al. 1985). In total, 121 objects were 
selected by these criteria.@As a next step, 47 objects were further selected according to 
the following criteria, 6)  previously unknown optical Miras candidates according to the 
SIMBAD database, 7) light variation of the optical(I) counterpart larger than 0.2 mag, 
and freedom from nebulosity (Jiang et al. 1996). The positions of the selected stars 
are known to better than 1 arcsec and the finding charts are given by Jiang et al.(1996). 

The data reduction was performed in a standard way by using the DIGIPHOT and ASTUTIL 
packages inside the IRAF software package. All the images were de-biased and then 
flat-fielded with normalized dome flat images. In each image field, 5 comparison stars 
for photometric calibration were chosen, within a few arc minutes around the object. We 
tried to pick up comparison stars having colors as similar as possible to those of the 
objects in order to reduce the differences between Kiso, Osaka and standard photometric 
systems. Figure 1 shows an example of object field images, where C1, C2 etc., denote 
the photometric comparison stars. Light variation of the objects at I--band can be 
easily recognizable in Figure 1. Differential photometry were performed between the 
comparison stars, and those with light variation larger than 0.1 mag were rejected for 
further use as photometric calibration. The magnitudes of comparison stars were obtained 
by reference to the equatorial standard stars (Landolt 1992). Only data obtained on 
photometric nights were adopted. The standardized magnitudes of the program objects 
were then determined by differential photometry with respect to the comparison stars. 
Figure 2 displays the photometric results of the 47 objects. The normal observational 
spans are 1500 days and the number of data points for each source is about 20. The 
mean I magnitudes, the amplitudes of variation, the number of data points, and the 
observational time baselines are listed in Table 2.\par
\vspace{1pc}\par

\begin{fv}{1}{18pc}%
{Examples of I--band images of object fields. Upper panels show the field of IRAS 00336+6744. 
Lower panels are the field of IRAS 04402+3426. Photometric comparison stars are shown by symbols 
as C1 -- C5. Images were taken by the 51cm Cassegrain telescope, at Osaka Kyoiku University.}
\end{fv}

\begin{fv}{2}{18pc}%
{The I--band light curves of the 47 color-selected IRAS PSC sources (names labeled at 
the lower-light corner). The dots represent the observational points. The solid curve in the 
diagram is a third-order Fourier fit to the data points, with the period 
determined by the PDM method. The period is shown at upper-right corner when it is well defined.}
\end{fv}

\section
{Period and Distance Determinations}

\subsection
{Period Determination}
The period was determined using the Phase Dispersion Minimization (PDM) method 
(Stellingwerf 1978). The PDM method is well suited to the case of nonsinusoidal 
time variation covered by only a few irregularly spaced observations. Because our 
observation is interrupted by the summer season and sometimes also by bad 
weather during the season when the objects are visible in the night sky, the 
photometric points are not regularly spaced in time. The type of time variation 
is presumed known.  The PDM method is applied to determine the period of the objects 
observed. In the PDM method, the data are divided by the trial period and ordered in 
according to the period. Then the average light curve (in a small span) and the 
dispersion of the data around it are calculated. The minimum value of the dispersion 
is searched over a range of trial periods. The results of PDM analysis of the 
observational data are shown in column 6, 7 and 8 of Table 2, which contain the period, 
the uncertainty in the period and the date of the light maximum respectively, for each 
object. As can be seen from Table 2, some of the period are determined well and some 
are not, which mainly depends on the parameter $theta$ (normalized dispersion) during 
the PDM analysis. As an example, Figure 3 shows the behavior of $theta$ in both cases 
of well-defined and uncertain periods. The horizontal axis is the trial period and the 
vertical axis is the normalized phase dispersion at the trial period. The minima of the 
phase dispersion indicate candidates for the real period. When the primary minimum of 
$theta$ is smaller than 0.5 and twice or more deeper than the secondary minimum, and 
the corresponding period is shorter than 700 days, the results are considered to be 
the true period. In the case of the periods over 700 days, which is roughly a half 
the observational span, a large uncertainty is involved because the observational 
time span is too short. In the $theta$ diagram, the line profile near the minimum 
is taken to be parabolic. If we extrapolate this parabola to $theta = 1$, the half width 
($\delta P$) for the line is given approximately by (Stellingwerf 1978);
\begin{equation}
   \Delta P \simeq \frac{P^2}{2T}
\end{equation}		
where $P$ is the trial period at the primary minimum, and $T$ is the time baseline of 
observation. In this work, $\Delta P$ is taken to be the uncertainty in the period and 
is the figure given in column 7 of Table 2. From this analysis, the periods of 18 objects 
could be determined while for the other 29, the conditions for a good solution to the 
period analysis were not satisfied and their period were not determined. The reasons 
for the indeterminacy were a small amplitude(00589+5743), insufficient data 
points(00534+6031, 02470+5536, 21449+4950 etc.), irregular time variation(03557+4404), 
and too short or too long period(00459+6749, 03096+5936 etc.). 

Figure 4 shows the histogram of the light variation periods of the objects. The upper 
panel shows the histogram for stars with well-determined period. The lower panel is 
for all sources including uncertain periods. The peak of period distribution is at 
450 -- 500 day in the histogram of the present sample. If we rely on the 
period--luminosity relation for Miras(Hughes \& Wood 1990; Feast et al 1989), a 
period 420 day corresponds to about 8000 $\LO$.  As far as the present sample is 
concerned, the mean value of periods tends to be slightly larger than 420 days. 
Almost all solutions with a short period($\sim$ 100 - 200 days) in the lower panel 
are false, because of insufficient time resolution. Among these short period sources, 
however, as far as 00336+6744 is concerned, the solution satisfies the period detection 
criteria. Some sources with the period longer than 700 day can be seen in the lower panel 
of Figure 4. These long-period sources were counted as uncertain, because of 
insufficient observation time baseline for calculation by the PDM method. From Figure 4, 
it can be seen that the detection rate of SiO masers is the highest for sources at the 
period 400 - 500 days. Whitelock et al. (1991) pointed out the absence of periods around 
350 day, in their work in the direction 
of the Bulge. In our result concerning the direction of the outer disk, the absence Miras 
around 350 day can also be seen as well (see Figure 4 and 5).

Figure 5 shows a period--color ($C_{12}$) diagram. No clear correlation between period 
and IRAS color can be seen. The data points in Figure 5 are divided into 2 groups, 
namely, a short-period group around 200 days and a long-period group around 500 days. 
It is possible to make the interpretation that the short-period group sources pulsate in the 
first-overtone mode which is dominant in Miras. Unfortunately, however, most of short 
period sources did not satisfy the period detection criteria because of insufficient 
time resolution. High time-resolution data are needed to determine the mode of  
pulsation. The pulsational mode of Miras is a highly controversial issue and remains
uncertain (Wood and Sebo 1996).\par
\vspace{1pc}\par          
       
\begin{fv}{3}{18pc}%
{Examples of theta diagram of the PDM analysis. The left panels show the sources 
with well-determined periods, the right panels show sources with undetermined 
period.}
\end{fv}

\begin{fv}{4}{18pc}%
{Histogram of period. The upper panel shows the histogram of period for the sources 
with well-determined period. The lower panel shows the histogram for all sources,  
i.e., it includes the sources with uncertain period.}
\end{fv}

\begin{fv}{5}{18pc}%
{Plot of period against IRAS color ($C_{12}$). Small symbols indicate the sources 
with uncertain period, that is, these sources did not satisfy the period determination 
criteria. (see text)}
\end{fv}

\subsection
{Luminosity Calculations}
The luminosity can be computed from the observed period and the period--luminosity 
relation. The luminosity of each individual source has been determined from two 
slightly different period--luminosity relations. These are period to I magnitude 
relation (Feast et al 1989; Reid et al 1988), and the period to bolometric magnitude 
relation (Hughes \& Wood 1990; Feast et al 1989). Both of these relations were 
determined from observations of Miras in the LMC. The period--luminosity relation of 
Miras depends on the C/O abundance ratio within the envelope (Feast et al 1989). Whether 
an object is C-rich or O-rich was judged from its IRAS Low Resolution Spectrum 
(Olnon et al. 1986), by detection of the SiO/OH/H$_{2}$O maser lines (Jiang et al. 
1997), by the HCN maser line (Loup et al. 1993), or by its optical spectrum (Jiang 1997). 
The assignment to C-rich or O-rich type is shown in column 8 of Table 3. Most of 
the sources with well-determined periods are found to have O-rich circumstellar envelope 
except for four sources: 02272+6327, 05273+2019, 05452+2001 and 05484+3521. The C/O ratio 
of 02272+6327 is unknown, while the latter three have C-rich envelopes and their 
absolute magnitude could not be calculated from their I--band magnitudes because the 
period to I--band magnitude relation for C-rich Miras was not yet well determined. We 
adopted the following period--luminosity relations (Hughes \& Wood 1990; Feast et al. 
1998; Reid et al. 1988) to calculate the luminosities of the objects:
\begin{equation}
   \langle I \rangle_{abs} = -1.23 \log P -1.51 \qquad (\sigma = 0.42)
\end{equation}		
for O-rich stars,
\begin{equation}
   \langle M_{bol} \rangle_{abs} = -2.91 \log P + 2.59 \qquad (\sigma = 0.32)
\end{equation}		
for O-rich stars with $P < 450$ day,
\begin{equation}
   \langle M_{bol} \rangle_{abs} = -7.76 \log P + 15.40 \qquad (\sigma = 0.38)
\end{equation}		
for O-rich stars with $P > 450$ day,
\begin{equation}
   \langle M_{bol} \rangle_{abs} = -1.86 \log P + 0.26 \qquad (\sigma = 0.13)
\end{equation}		
for C-rich stars, where $\langle I \rangle_{abs}$ and $\langle M_{bol} \rangle_{abs}$ 
are average 
I--band absolute magnitude and bolometric absolute magnitude, respectively, and 
$\sigma$ is the 
standard deviation of the absolute magnitudes. 
The luminosity of each star can also be obtained 
from $\langle M_{bol} \rangle_{abs}$ by the following equation (Zombeck 1982).
\begin{equation}
   \langle M_{bol} \rangle_{abs} - 4.72 = -2.5 \log ({L}/{\LO})
\end{equation}		
The calculated absolute I magnitude and absolute bolometric magnitude of the 
objects from both relations are shown in column 2 and 3 of Table 3. The absolute 
I-magnitude are left blank due to the lack of period--I-band absolute magnitude 
relation. 

It is known that the slope of the period--luminosity relation of Miras slightly 
changes at the period of about 450 days as indicates in Equations (3) and (4) 
(Hughes \& Wood 1990). These relations were obtained from near-infrared observations; 
most of the energy of the AGB stars is radiated at infrared wavelengths. In contrast, 
Equations (2) and (5) do not involve this slope change at longer periods. This is 
because the observations used in deriving equations (2) and (5) were made by 
photographic plates (Reid et al. 1988) which were not sensitive enough to the 
red stars with long periods, i.e.,  the number of objects with longer periods 
was insufficient to reveal the change of slope.In addition, we should be note that 
I-band period-luminosity relation is likely to be somewhat affected by metallicity 
difference between LMC and our Galaxy (Feast 1996).

\subsection
{Distance Calculations and Results}
Distance is an important fundamental parameter and difficult to determine accurately. 
Taking the known absolute and apparent magnitude, we can estimate the distance of the 
objects relatively accurately. The distances were calculated by two different methods, 
one using I magnitude, and the other using the bolometric magnitude and IRAS 12/25 
micron ratio.

\subsubsection
{METHOD 1}
The distance of each source is determined from the difference between the apparent 
and absolute I--band magnitudes. At the wavelength of the I--band, interstellar 
extinction corrections must be applied;
\begin{equation}
   m - M - A(D) = 5 \log [D/(10pc)]
\end{equation}		
where $A(D)$ means the interstellar extinction up to the distance $D$ and $m$ and $M$ are the 
apparent and absolute magnitude, respectively. Here, two different extinction models are used to 
define the functional form of $A(D)$. The first model assumes a homogeneus dust distribution 
on the line of sight (MODEL 1), i.e., the extinction increases with distance; 
\begin{equation}
   A(D) = a(l) D
\end{equation}		
where $a(l)$, the extinction per unit distance, depends on galactic longitude $l$. 
It is taken from the optical observations of nearby open clusters within 1 kpc 
(Chen et al. 1998) and  computed for every 10 degrees in galactic longitude. It 
varies in the range 0.52 - 1.17 mag kpc$^{-1}$. The known extinction ratio $A_{I}/A_{V}$ 
(He et al. 1995) is used to convert the extinction in the V--band to that in the I--band. 

An alternative model assumes an exponential dust distribution (MODEL 2),
\begin{equation}
   \alpha _{I}(D) = 0.776 (mag/kpc) \exp [- \frac{r - r_{0}}{r_{h}} - \frac{|z| - |z_{0}|}{z_{h}}]
\end{equation}		
where $r$ and $z$ are the galactocentric distance and height from the galactic plane, 
$r_{h}$ and $z_{h}$ are the scale height in radial and perpendicular directions, and $r_{0}$ 
and $z_{0}$ are the galactocentric distance and height from the galactic plane at the 
Sun, respectively. We adopt the values, $r_{h} = 3.4 kpc$, $z_{h} = 40 pc$, $r_{0} = 
8.5 kpc$, $z_{0} = -14 pc$ (Unavane et al. 1998, Kerr \& Lynden-Bell 1986). The Extinction 
$A(D)$ is the integration over distance as
\begin{equation}
   A(D) = \int_0^D \alpha _{I}(D')\,dD'
\end{equation}		
The results of distance calculation using Equation (7), (8), (9) and (10), are given in 
Table 4. In Table 4, Columns 2 and 4 show the calculated distances and columns 3 and 5 
show the interstellar extinction up to the calculated distance, for models 1 and 2, 
respectively.  The symbol $^{*}$ in Table 4 denotes distances further than 10 kpc. 
These apparently large distance may result from, for instance, presence of dust clouds 
in the line of sight, or from large circumstellar extinction at the star itself. 
However, no clear evidence of dust clouds in the line of sight of these sources appears 
in the SIMBAD data base. Figure 6 shows the positions of the observed sources projected 
onto the Galactic plane, using the distances calculated by METHOD 1. \par
\vspace{1pc}\par

\begin{fv}{6}{18pc}%
{Positions of the observed sources projected onto the Galactic plane. The position
of the Sun is at the origin. The dashed lines indicate circles around the
galactic centre with radii from 8 to 18 $kpc$ in every 2 $kpc$ intervals. }  
\end{fv}

\subsubsection
{METHOD 2}
In this method, the distance is calculated from the IRAS 12$\mu m$ flux density and the 
luminosity (which is calculated from the absolute bolometric magnitude).  
The bolometric flux is calculated from the 
IRAS 12$\mu m$ flux data using a bolometric correction. The relation between  
the bolometric flux and the IRAS 12$\mu m$ flux is given by
\begin{equation}
   F_{bol} = 2.5 \times 10^{-10} (erg/s) (BC)_{12} (F_{12}/Jy)
\end{equation}		
where $F_{bol}$, $F_{12}$ and $(BC)_{12}$ are the bolometric flux, the IRAS 12$\mu m$ 
flux density, and the bolometric correction, respectively. Approximate formula for 
$(BC)_{12}$ is given by (van der Veen \& Breukers 1989)
\begin{equation}
   (BC)_{12} = 0.7 + 2.9\,e^{-7.5 \times C_{12}} + 0.9\,e^{1.75 \times C_{12}}
\end{equation}		
The distance can then be calculated as
\begin{equation}
   D_{bol} = \sqrt{L/(4\pi \,F_{bol})},
\end{equation}		
where L has been already given in Equation 6. In this distance calculation, we 
rely on the assumption that all the energy of a source is radiated in the mid-infrared 
wavelength region. Interstellar extinction is negligible at these wavelengths and  
therefore extinction corrections are not applied. The resultant distance and $(BC)_{12}$ 
value are shown in column 6 and 7 of Table 4. Figure 7 shows the projected positions of 
our sample on the galactic plane using METHOD 2.\par
\vspace{1pc}\par

\begin{fv}{7}{18pc}%
{The same as Figure 6 but using the distances calculated by METHOD 2. }  
\end{fv}

\section{Discussion}
In the previous work (Jiang et al. 1997), distances to the objects were determined 
on the assumption of uniform luminosity (8000 $\LO$). Positions projected onto 
galactic plane using the constant luminosity (8000 $\LO$) are plotted in Figure 7 
together with those found using METHOD 2. No clear systematic difference can be seen 
between METHOD 2 and Jiang et al. positions.

If we assume that the stars are in circular orbits around the galactic center, then 
as we know their distances, the radial velocities can be converted to circular velocities. 
From these we can obtain the rotation curve. Radial velocities are known from previous 
SiO maser observations for 23 objects in the present sample (Jiang et al. 1997; Nakashima 
et al. 1999 to be published); they are shown in column 7 of Table 3. We plot the rotation 
curve of the outer galactic disk in Figure 8. The results of least-square fit are;
\begin{equation}
     \Theta = 220\,km\,s^{-1} - 9.1(\pm 7.3)km\,s^{-1}\,kpc^{-1} \times (R - R_{0})\,kpc   
\end{equation}		
for METHOD 1 (MODEL 1),
\begin{equation}
     \Theta = 220\,km\,s^{-1} - 15.4(\pm 7.7)km\,s^{-1}\,kpc^{-1} \times (R - R_{0})\,kpc   
\end{equation}		
for MEHOD 2, where $R_{0}$ equals 8.5 kpc. The slope of the rotation curves found in this study 
tend to be flatter than those obtained previously (Jiang et al. 1996). From the rotation 
curves, the Oort's constants are computed to be: $A = 17.5(\pm 3.7)\,km\,s^{-1}\,kpc^{-1}$, 
$B = -8.4(\pm 3.7)\,km\,s^{-1}\,kpc^{-1}$ (for METHOD 1), 
$A = 20.6(\pm 3.9)\,km\,s^{-1}\,kpc^{-1}$, $B = -5.2(\pm 3.9)\,km\,s^{-1}\,kpc^{-1}$ 
(for METHOD 2). These value differs slightly from those found in previous investigations, 
for instance studies of OB star and Cepheid (Frink et al. 1996), and IAU standard value
($A = 15\,km\,s^{-1}\,kpc^{-1}$, 
$B = -10\,km\,s^{-1}\,kpc^{-1}$) 
(Kerr \& Lynden-Bell 1986). Unfortunately, it is difficult to discuss the difference 
in detail, because of the small size of the present sample.  
 
The distances calculated by METHOD 1 contain large errors because of the interstellar 
extinction corrections. Judging from the distribution of nearby interstellar clouds 
(Chen et al. 1998), the errors involved in distance calculation are estimated as a few 
$kpc$. On the other hand, interstellar extinction corrections are not needed in the case 
of METHOD 2.  Therefore METHOD 2 should give a better distance than METHOD 1. Errors in 
the case of METHOD 2  are estimated at 30 \% from the uncertainties in the 
period--luminosity relation and observed magnitudes.

In the distance calculation of METHOD 2, we rely on the assumption that all energy of 
a source is radiated in the mid-infrared wavelength region. However, the luminosities 
which are used in this method may be slightly higher than the mid-infrared luminosities 
by themselves, because part of the energy is also radiated at near-infrared and visible wavelengths. In order to obtain more accurate luminosity distances, we need to know the 
spectral energy distribution of each source. 

According to the period--luminosity relation, objects at 8000 $\LO$ have a pulsation 
period of 420 days. A constant value of 8000 $\LO$ was therefore used to determine the 
luminosity distance in our previous work. On the other hand, the mean pulsation period 
of objects found in this work is about 500 day, which corresponds to 10000 $\LO$.\par
\vspace{1pc}\par

\begin{fv}{8}{18pc}%
{The rotation curve of the outer disk, for methods 1(MODEL 1) and 2. 
The solid lines are fitted to the points corresponding to objects with well-determined
periods.}  
\end{fv}

\section{Conclusion}
In this paper, we have reported the results of photometric observations to determine 
the distances of the AGB stars using the period luminosity relation. Periods were well 
determined for about 18 sources of 47 color-selected AGB candidates, distributed in the 
outer disk of the Galaxy. The mean value of the pulsation period is about 500 day, 
which corresponds to 10000 $\LO$ according to the period--luminosity relation. This is 
slightly more luminous than the value of 8000 $\LO$ which was used to determine 
luminosity distance in previous works. Distances were calculated by two different 
methods, viz. METHOD 1 and 2. The results from METHOD 1 contain large uncertainty, 
because of the interstellar extinction corrections. METHOD 2 is more reliable than 
METHOD 1, because interstellar extinction corrections are not needed. 

Characteristics of individual objects are also determined; IRAS 00336+6744 is the only 
object which satisfies the criteria of period determination amongst the objects with 
PDM solution shorter than 300 days; IRAS 04209+4800 potentially has a pulsation period 
longer than 800 day. But period determination criteria were not always satisfied in the 
present analysis, because of the insufficient time span of the observations. \par
\vspace{1pc}\par
The authors thank the staff of Kiso observatory and Astronomical Institute, Osaka 
Kyoiku Univ., for their help during the long term observations. They also thank I.S. Glass 
for reading the manuscript and the comments. One of the authors 
(B.W.J.) received a Japanese government scholarship for foreign students. This 
research was supported by the Grant in-aid for Scientific Research (C) No.08640337 
and 10640238 of Ministry of Education, Science, Sports and Culture.

\begin{table*}
\renewcommand{\baselinestretch}{1.0}
\small
\begin{center}
Table~1.\hspace{4pt}Observation log of the photometric observations. The IRAS name and Julian Dates are listed. \\
\end{center}
\vspace{6pt}
\begin{tabular*}{\textwidth}{@{\hspace{\tabcolsep}
\extracolsep{\fill}}p{5pc}cccccccccccccc}
\hline\hline\\[-6pt]
IRAS NAME & \multicolumn{13}{c}{Obsevation Date (Julian Date - 2400000.0)} \\
[4pt]\hline\\[-6pt]
{\footnotesize 00336+6744} & {\footnotesize 49661} & {\footnotesize 49687} & {\footnotesize 49700} & {\footnotesize 49726} & {\footnotesize 49741} & {\footnotesize 49759} & {\footnotesize 49913} &
 {\footnotesize 49949} & {\footnotesize 49978} & {\footnotesize 50040} & {\footnotesize 50053} & {\footnotesize 50109} & {\footnotesize 50466}  \\
{\footnotesize 00459+6749} & {\footnotesize 49661} & {\footnotesize 49687} & {\footnotesize 49726} & {\footnotesize 49741} & {\footnotesize 49759} & {\footnotesize 49913} & {\footnotesize 49978} & {\footnotesize 50040} & {\footnotesize 50053} & {\footnotesize 50109} & {\footnotesize 50466} & {\footnotesize 50662} & {\footnotesize 50676} \\
{\footnotesize 00534+6031} & {\footnotesize 49610} & {\footnotesize 49627} & {\footnotesize 50662} & {\footnotesize 50676} & {\footnotesize 50711} & {\footnotesize 50728} & {\footnotesize 50739} & {\footnotesize 50760} & {\footnotesize 50772} & {\footnotesize 50785} & {\footnotesize 50830} & {\footnotesize 51073} & {\footnotesize } \\
{\footnotesize 00589+5743} & {\footnotesize 49607} & {\footnotesize 49627} & {\footnotesize 40659} & {\footnotesize 49686} & {\footnotesize 49700} & {\footnotesize 49726} & {\footnotesize 49741} & {\footnotesize 49759} & {\footnotesize 49948} & {\footnotesize 49978} & {\footnotesize 50053} & {\footnotesize 50109} & {\footnotesize 50466} \\
{\footnotesize 02117+5559} & {\footnotesize 49607} & {\footnotesize 49697} & {\footnotesize 49726} & {\footnotesize 49742} & {\footnotesize 49759} & {\footnotesize 49798} & {\footnotesize 49913} & {\footnotesize 49949} & {\footnotesize 49978} & {\footnotesize 50039} & {\footnotesize 50053} & {\footnotesize 50110} & {\footnotesize 50320} \\
{\footnotesize 02272+6327} & {\footnotesize 49607} & {\footnotesize 49627} & {\footnotesize 49660} & {\footnotesize 49686} & {\footnotesize 49700} & {\footnotesize 49726} & {\footnotesize 49742} & {\footnotesize 49759} & {\footnotesize 49798} & {\footnotesize 49948} & {\footnotesize 49978} & {\footnotesize 50039} & {\footnotesize 50053} \\
{\footnotesize 02433+6345} & {\footnotesize 49565} & {\footnotesize 49606} & {\footnotesize 49697} & {\footnotesize 49725} & {\footnotesize 49742} & {\footnotesize 49759} & {\footnotesize 49798} & {\footnotesize 49948} & {\footnotesize 49978} & {\footnotesize 50039} & {\footnotesize 50053} & {\footnotesize 50110} & {\footnotesize 50320} \\
{\footnotesize 02470+5536} & {\footnotesize 49606} & {\footnotesize 49627} & {\footnotesize 50662} & {\footnotesize 50676} & {\footnotesize 50728} & {\footnotesize 50739} & {\footnotesize 50763} & {\footnotesize 50788} & {\footnotesize 50833} & {\footnotesize 51073} & {\footnotesize } & {\footnotesize } & {\footnotesize } \\
{\footnotesize 03022+5409} & {\footnotesize 49610} & {\footnotesize 49627} & {\footnotesize 49697} & {\footnotesize 49725} & {\footnotesize 49742} & {\footnotesize 49759} & {\footnotesize 49798} & {\footnotesize 49948} & {\footnotesize 49978} & {\footnotesize 50039} & {\footnotesize 50053} & {\footnotesize 50110} & {\footnotesize 50152} \\
{\footnotesize 03096+5936} & {\footnotesize 49607} & {\footnotesize 49627} & {\footnotesize 49658} & {\footnotesize 49686} & {\footnotesize 49743} & {\footnotesize 49759} & {\footnotesize 49798} & {\footnotesize 49948} & {\footnotesize 50039} & {\footnotesize 50053} & {\footnotesize 50110} & {\footnotesize 50152} & {\footnotesize 50466} \\
{\footnotesize 03192+5642} & {\footnotesize 49607} & {\footnotesize 49627} & {\footnotesize 49658} & {\footnotesize 49686} & {\footnotesize 49700} & {\footnotesize 49742} & {\footnotesize 49759} & {\footnotesize 49798} & {\footnotesize 49949} & {\footnotesize 50039} & {\footnotesize 50053} & {\footnotesize 50110} & {\footnotesize 50152} \\
{\footnotesize 03238+6034} & {\footnotesize 49627} & {\footnotesize 49658} & {\footnotesize 49686} & {\footnotesize 49700} & {\footnotesize 49726} & {\footnotesize 49742} & {\footnotesize 49759} & {\footnotesize 49798} & {\footnotesize 49949} & {\footnotesize 50039} & {\footnotesize 50053} & {\footnotesize 50110} & {\footnotesize 50117} \\
{\footnotesize 03371+4932} & {\footnotesize 49627} & {\footnotesize 49658} & {\footnotesize 49726} & {\footnotesize 49742} & {\footnotesize 49759} & {\footnotesize 49799} & {\footnotesize 49949} & {\footnotesize 50039} & {\footnotesize 50053} & {\footnotesize 50110} & {\footnotesize 50466} & {\footnotesize 50830} & {\footnotesize 50891} \\
{\footnotesize 03469+5833} & {\footnotesize 49660} & {\footnotesize 49686} & {\footnotesize 49700} & {\footnotesize 49742} & {\footnotesize 49759} & {\footnotesize 49799} & {\footnotesize 49948} & {\footnotesize 50039} & {\footnotesize 50053} & {\footnotesize 50110} & {\footnotesize 50466} & {\footnotesize 50672} & {\footnotesize 50694} \\
{\footnotesize 03525+5711} & {\footnotesize 49627} & {\footnotesize 49658} & {\footnotesize 49686} & {\footnotesize 49700} & {\footnotesize 49725} & {\footnotesize 49726} & {\footnotesize 49742} & {\footnotesize 49759} & {\footnotesize 49799} & {\footnotesize 49948} & {\footnotesize 50039} & {\footnotesize 50053} & {\footnotesize 50110} \\
{\footnotesize 03557+4404} & {\footnotesize 49627} & {\footnotesize 49697} & {\footnotesize 49726} & {\footnotesize 49743} & {\footnotesize 49760} & {\footnotesize 49799} & {\footnotesize 49948} & {\footnotesize 49949} & {\footnotesize 50039} & {\footnotesize 50053} & {\footnotesize 50110} & {\footnotesize 50466} & {\footnotesize 50694} \\
{\footnotesize 04085+5347} & {\footnotesize 49627} & {\footnotesize 49658} & {\footnotesize 49686} & {\footnotesize 49700} & {\footnotesize 49726} & {\footnotesize 49741} & {\footnotesize 49757} & {\footnotesize 49797} & {\footnotesize 49948} & {\footnotesize 50039} & {\footnotesize 50053} & {\footnotesize 50110} & {\footnotesize 50152} \\
{\footnotesize 04209+4800} & {\footnotesize 49660} & {\footnotesize 49687} & {\footnotesize 49697} & {\footnotesize 49726} & {\footnotesize 49741} & {\footnotesize 49757} & {\footnotesize 49797} & {\footnotesize 49948} & {\footnotesize 50039} & {\footnotesize 50053} & {\footnotesize 50110} & {\footnotesize 50119} & {\footnotesize 50152} \\
{\footnotesize 04402+3426} & {\footnotesize 49661} & {\footnotesize 49686} & {\footnotesize 49700} & {\footnotesize 49726} & {\footnotesize 49741} & {\footnotesize 49757} & {\footnotesize 49798} & {\footnotesize 49948} & {\footnotesize 50039} & {\footnotesize 50053} & {\footnotesize 50109} & {\footnotesize 50152} & {\footnotesize 50468} \\
{\footnotesize 04470+3002} & {\footnotesize 49660} & {\footnotesize 49686} & {\footnotesize 49726} & {\footnotesize 49741} & {\footnotesize 49757} & {\footnotesize 49799} & {\footnotesize 49948} & {\footnotesize 50039} & {\footnotesize 50053} & {\footnotesize 50109} & {\footnotesize 50152} & {\footnotesize 50468} & {\footnotesize 50694} \\
{\footnotesize 05091+4639} & {\footnotesize 49660} & {\footnotesize 49686} & {\footnotesize 49700} & {\footnotesize 49725} & {\footnotesize 49741} & {\footnotesize 49757} & {\footnotesize 49797} & {\footnotesize 49948} & {\footnotesize 50039} & {\footnotesize 50053} & {\footnotesize 50109} & {\footnotesize 50152} & {\footnotesize 50468} \\
{\footnotesize 05146+2521} & {\footnotesize 49660} & {\footnotesize 49686} & {\footnotesize 49725} & {\footnotesize 49743} & {\footnotesize 49758} & {\footnotesize 49797} & {\footnotesize 49948} & {\footnotesize 50039} & {\footnotesize 50053} & {\footnotesize 50109} & {\footnotesize 50119} & {\footnotesize 50152} & {\footnotesize 50735} \\
{\footnotesize 05204+3227} & {\footnotesize 49661} & {\footnotesize 49686} & {\footnotesize 49743} & {\footnotesize 49758} & {\footnotesize 49797} & {\footnotesize 49949} & {\footnotesize 50039} & {\footnotesize 50053} & {\footnotesize 50109} & {\footnotesize 50152} & {\footnotesize 50468} & {\footnotesize 50735} & {\footnotesize 50742} \\
{\footnotesize 05273+2019} & {\footnotesize 49687} & {\footnotesize 49725} & {\footnotesize 49743} & {\footnotesize 49758} & {\footnotesize 49797} & {\footnotesize 49949} & {\footnotesize 50039} & {\footnotesize 50053} & {\footnotesize 50109} & {\footnotesize 50152} & {\footnotesize 50468} & {\footnotesize 50735} & {\footnotesize 50742} \\
{\footnotesize 05405+3240} & {\footnotesize 49661} & {\footnotesize 49686} & {\footnotesize 49758} & {\footnotesize 49797} & {\footnotesize 49949} & {\footnotesize 50039} & {\footnotesize 50053} & {\footnotesize 50109} & {\footnotesize 50152} & {\footnotesize 50468} & {\footnotesize 50735} & {\footnotesize 50742} & {\footnotesize 50785} \\
{\footnotesize 05423+2905} & {\footnotesize 49660} & {\footnotesize 49686} & {\footnotesize 49725} & {\footnotesize 49741} & {\footnotesize 49757} & {\footnotesize 49799} & {\footnotesize 49949} & {\footnotesize 50039} & {\footnotesize 50053} & {\footnotesize 50109} & {\footnotesize 50152} & {\footnotesize 50468} & {\footnotesize 50735} \\
{\footnotesize 05452+2001} & {\footnotesize 49687} & {\footnotesize 49726} & {\footnotesize 49743} & {\footnotesize 49758} & {\footnotesize 49799} & {\footnotesize 49949} & {\footnotesize 50039} & {\footnotesize 50053} & {\footnotesize 50109} & {\footnotesize 50152} & {\footnotesize 50468} & {\footnotesize 50735} & {\footnotesize 50785} \\
{\footnotesize 05484+3521} & {\footnotesize 49661} & {\footnotesize 49686} & {\footnotesize 49725} & {\footnotesize 49743} & {\footnotesize 49758} & {\footnotesize 49797} & {\footnotesize 49949} & {\footnotesize 50039} & {\footnotesize 50053} & {\footnotesize 50109} & {\footnotesize 50152} & {\footnotesize 50468} & {\footnotesize 50735} \\
{\footnotesize 05552+1720} & {\footnotesize 49687} & {\footnotesize 49726} & {\footnotesize 49743} & {\footnotesize 49759} & {\footnotesize 49760} & {\footnotesize 49799} & {\footnotesize 49949} & {\footnotesize 50039} & {\footnotesize 50053} & {\footnotesize 50109} & {\footnotesize 50152} & {\footnotesize 50468} & {\footnotesize 50735} \\
{\footnotesize 06170+3523} & {\footnotesize 49660} & {\footnotesize 49686} & {\footnotesize 49725} & {\footnotesize 49741} & {\footnotesize 49757} & {\footnotesize 49797} & {\footnotesize 49950} & {\footnotesize 50039} & {\footnotesize 50053} & {\footnotesize 50109} & {\footnotesize 50152} & {\footnotesize 50468} & {\footnotesize 50570} \\
{\footnotesize 06447+0817} & {\footnotesize 49686} & {\footnotesize 49726} & {\footnotesize 49743} & {\footnotesize 49758} & {\footnotesize 49798} & {\footnotesize 50039} & {\footnotesize 50053} & {\footnotesize 50109} & {\footnotesize 50110} & {\footnotesize 50152} & {\footnotesize 50468} & {\footnotesize 50570} & {\footnotesize 50735} \\
{\footnotesize 06448+1639} & {\footnotesize 49686} & {\footnotesize 49726} & {\footnotesize 49743} & {\footnotesize 49798} & {\footnotesize 50039} & {\footnotesize 50570} & {\footnotesize 50735} & {\footnotesize 50742} & {\footnotesize 50785} & {\footnotesize 50830} & {\footnotesize 50879} & {\footnotesize 51105} & {\footnotesize 51130} \\
{\footnotesize 20523+5302} & {\footnotesize 49565} & {\footnotesize 49606} & {\footnotesize 49697} & {\footnotesize 49726} & {\footnotesize 49741} & {\footnotesize 49948} & {\footnotesize 49978} & {\footnotesize 50039} & {\footnotesize 50053} & {\footnotesize 50468} & {\footnotesize 50662} & {\footnotesize 50676} & {\footnotesize 50702} \\
{\footnotesize 20532+5554} & {\footnotesize 49562} & {\footnotesize 49607} & {\footnotesize 49658} & {\footnotesize 49687} & {\footnotesize 49697} & {\footnotesize 49726} & {\footnotesize 49742} & {\footnotesize 49913} & {\footnotesize 49949} & {\footnotesize 49978} & {\footnotesize 50039} & {\footnotesize 50053} & {\footnotesize 50676} \\
{\footnotesize 21086+5238} & {\footnotesize 49562} & {\footnotesize 49606} & {\footnotesize 49627} & {\footnotesize 49658} & {\footnotesize 49690} & {\footnotesize 49726} & {\footnotesize 49742} & {\footnotesize 49913} & {\footnotesize 49949} & {\footnotesize 49978} & {\footnotesize 50040} & {\footnotesize 50054} & {\footnotesize 50468} \\
{\footnotesize 21216+5536} & {\footnotesize 49565} & {\footnotesize 49606} & {\footnotesize 49627} & {\footnotesize 49658} & {\footnotesize 49690} & {\footnotesize 49726} & {\footnotesize 49742} & {\footnotesize 49913} & {\footnotesize 49949} & {\footnotesize 49978} & {\footnotesize 50040} & {\footnotesize 50054} & {\footnotesize 50468} \\
{\footnotesize 21377+5042} & {\footnotesize 49562} & {\footnotesize 49707} & {\footnotesize 49726} & {\footnotesize 49743} & {\footnotesize 49913} & {\footnotesize 49948} & {\footnotesize 49978} & {\footnotesize 50040} & {\footnotesize 50054} & {\footnotesize 50468} & {\footnotesize 50676} & {\footnotesize 50702} & {\footnotesize 50710} \\
{\footnotesize 21415+5025} & {\footnotesize 49562} & {\footnotesize 49607} & {\footnotesize 49627} & {\footnotesize 49658} & {\footnotesize 49690} & {\footnotesize 49726} & {\footnotesize 49743} & {\footnotesize 49913} & {\footnotesize 49949} & {\footnotesize 49978} & {\footnotesize 50040} & {\footnotesize 50054} & {\footnotesize 50468} \\
{\footnotesize 21449+4950} & {\footnotesize 49562} & {\footnotesize 49948} & {\footnotesize 50039} & {\footnotesize 50054} & {\footnotesize 50662} & {\footnotesize 50676} & {\footnotesize 50702} & {\footnotesize 50710} & {\footnotesize 50711} & {\footnotesize 50739} & {\footnotesize 50749} & {\footnotesize 51112} & {\footnotesize } \\
{\footnotesize 21453+5959} & {\footnotesize 49565} & {\footnotesize 49607} & {\footnotesize 49659} & {\footnotesize 49726} & {\footnotesize 49948} & {\footnotesize 50039} & {\footnotesize 50662} & {\footnotesize 50676} & {\footnotesize 50702} & {\footnotesize 50711} & {\footnotesize 50739} & {\footnotesize 50749} & {\footnotesize 50785} \\
{\footnotesize 21509+6234} & {\footnotesize 49565} & {\footnotesize 49627} & {\footnotesize 49659} & {\footnotesize 49690} & {\footnotesize 49708} & {\footnotesize 49726} & {\footnotesize 49741} & {\footnotesize 49760} & {\footnotesize 49913} & {\footnotesize 49949} & {\footnotesize 49978} & {\footnotesize 50039} & {\footnotesize 50054} \\
{\footnotesize 21563+5630} & {\footnotesize 49565} & {\footnotesize 49627} & {\footnotesize 49659} & {\footnotesize 49690} & {\footnotesize 49708} & {\footnotesize 49726} & {\footnotesize 49741} & {\footnotesize 49760} & {\footnotesize 49948} & {\footnotesize 50039} & {\footnotesize 50054} & {\footnotesize 50110} & {\footnotesize 50662} \\
{\footnotesize 22241+6005} & {\footnotesize 49687} & {\footnotesize 49725} & {\footnotesize 49743} & {\footnotesize 49758} & {\footnotesize 49797} & {\footnotesize 49949} & {\footnotesize 50039} & {\footnotesize 50053} & {\footnotesize 50109} & {\footnotesize 50152} & {\footnotesize 50468} & {\footnotesize 50735} & {\footnotesize 50742} \\
{\footnotesize 22394+5623} & {\footnotesize 49661} & {\footnotesize 49741} & {\footnotesize 49760} & {\footnotesize 49913} & {\footnotesize 49949} & {\footnotesize 50039} & {\footnotesize 50053} & {\footnotesize 50110} & {\footnotesize 50662} & {\footnotesize 50676} & {\footnotesize 50702} & {\footnotesize 50711} & {\footnotesize 50728} \\
{\footnotesize 22394+6930} & {\footnotesize 49661} & {\footnotesize 49726} & {\footnotesize 49741} & {\footnotesize 49760} & {\footnotesize 49913} & {\footnotesize 49949} & {\footnotesize 50040} & {\footnotesize 50053} & {\footnotesize 50110} & {\footnotesize 50468} & {\footnotesize 50662} & {\footnotesize 50676} & {\footnotesize 50702} \\
{\footnotesize 22466+6942} & {\footnotesize 49565} & {\footnotesize 49627} & {\footnotesize 49659} & {\footnotesize 49686} & {\footnotesize 49726} & {\footnotesize 49741} & {\footnotesize 49760} & {\footnotesize 49948} & {\footnotesize 50040} & {\footnotesize 50053} & {\footnotesize 50110} & {\footnotesize 50662} & {\footnotesize 50676} \\
{\footnotesize 23491+6243} & {\footnotesize 49562} & {\footnotesize 49690} & {\footnotesize 50039} & {\footnotesize 50053} & {\footnotesize 50110} & {\footnotesize 50662} & {\footnotesize 50676} & {\footnotesize 50702} & {\footnotesize 50711} & {\footnotesize 50749} & {\footnotesize 50788} & {\footnotesize 50833} & {\footnotesize 51116} \\
   
\\[4pt]
\hline
\end{tabular*}
 
\vspace{6pt}

\noindent
\end{table*}

\begin{table*}
\renewcommand{\baselinestretch}{1.0}
\small
\begin{center}
Table~1.\hspace{4pt}(continued) \\
\end{center}
\vspace{6pt}
\begin{tabular*}{\textwidth}{@{\hspace{\tabcolsep}
\extracolsep{\fill}}p{5pc}cccccccccccccc}
\hline\hline\\[-6pt]
IRAS NAME & \multicolumn{13}{c}{Obsevation Date (Julian Date - 2400000.0)} \\
[4pt]\hline\\[-6pt]
{\footnotesize 00336+6744} & {\footnotesize 50662} & {\footnotesize 50676} & {\footnotesize 50711} & {\footnotesize 50728} &  {\footnotesize 50739} & {\footnotesize 50749} & {\footnotesize 50772} & {\footnotesize 50788} & {\footnotesize 50830} &  {\footnotesize 51073} & {\footnotesize } &
 {\footnotesize } & {\footnotesize  } \\
{\footnotesize 00459+6749} & {\footnotesize 50711} & {\footnotesize 50728} & {\footnotesize 50739} &  {\footnotesize 50760} &  {\footnotesize 50772} & {\footnotesize 50785} & {\footnotesize 50833} &  {\footnotesize 51073} & {\footnotesize } &  {\footnotesize } & {\footnotesize } & 
 {\footnotesize } & {\footnotesize  } \\
{\footnotesize 00534+6031} & {\footnotesize } & {\footnotesize } & {\footnotesize } &  {\footnotesize } &  {\footnotesize } & {\footnotesize } & {\footnotesize } &  {\footnotesize } & {\footnotesize } &  {\footnotesize } & {\footnotesize } & 
 {\footnotesize } & {\footnotesize  } \\
{\footnotesize 00589+5743} & {\footnotesize 50662} & {\footnotesize 50676} & {\footnotesize 50711} &  {\footnotesize 50728} &  {\footnotesize 50739} & {\footnotesize 50760} & {\footnotesize 50788} &  {\footnotesize 50833} & {\footnotesize 51073} &  {\footnotesize } & {\footnotesize } & 
 {\footnotesize } & {\footnotesize  } \\
{\footnotesize 02117+5559} & {\footnotesize 50466} & {\footnotesize 50662} & {\footnotesize 50676} &  {\footnotesize 50760} &  {\footnotesize 50785} & {\footnotesize 50830} & {\footnotesize 50891} &  {\footnotesize 51073} & {\footnotesize } &  {\footnotesize } & {\footnotesize } & 
 {\footnotesize } & {\footnotesize  } \\
{\footnotesize 02272+6327} & {\footnotesize 50110} & {\footnotesize 50466} & {\footnotesize 50662} &  {\footnotesize 50676} &  {\footnotesize 50711} & {\footnotesize 50739} & {\footnotesize 50760} &  {\footnotesize 50788} & {\footnotesize 50833} &  {\footnotesize 51073} & {\footnotesize } & 
 {\footnotesize } & {\footnotesize  } \\
{\footnotesize 02433+6345} & {\footnotesize 50466} & {\footnotesize 50662} & {\footnotesize 50676} &  {\footnotesize 50711} &  {\footnotesize 50728} & {\footnotesize 50739} & {\footnotesize 50760} &  {\footnotesize 50788} & {\footnotesize } &  {\footnotesize } & {\footnotesize } & 
 {\footnotesize } & {\footnotesize  } \\
{\footnotesize 02470+5536} & {\footnotesize } & {\footnotesize } & {\footnotesize } &  {\footnotesize } &  {\footnotesize } & {\footnotesize } & {\footnotesize } &  {\footnotesize } & {\footnotesize } &  {\footnotesize } & {\footnotesize } & 
 {\footnotesize } & {\footnotesize  } \\
{\footnotesize 03022+5409} & {\footnotesize 50466} & {\footnotesize 50662} & {\footnotesize 50676} &  {\footnotesize 50711} &  {\footnotesize 50728} & {\footnotesize 50739} & {\footnotesize 50763} &  {\footnotesize 50785} & {\footnotesize 50830} &  {\footnotesize 50879} & {\footnotesize 50891} & 
 {\footnotesize 51073} & {\footnotesize  } \\
{\footnotesize 03096+5936} & {\footnotesize 50662} & {\footnotesize 50676} & {\footnotesize 50711} &  {\footnotesize 50735} &  {\footnotesize 50739} & {\footnotesize 50763} & {\footnotesize 50788} &  {\footnotesize 50833} & {\footnotesize 51073} &  {\footnotesize 51105} & {\footnotesize 51116} & 
 {\footnotesize } & {\footnotesize  } \\
{\footnotesize 03192+5642} & {\footnotesize 50466} & {\footnotesize 50672} & {\footnotesize 50694} &  {\footnotesize 50711} &  {\footnotesize 50735} & {\footnotesize 50742} & {\footnotesize 50763} &  {\footnotesize 50788} & {\footnotesize 50830} &  {\footnotesize 50891} & {\footnotesize 51116} & 
 {\footnotesize } & {\footnotesize  } \\
{\footnotesize 03238+6034} & {\footnotesize 50672} & {\footnotesize 50694} & {\footnotesize 50711} &  {\footnotesize 50742} &  {\footnotesize 50763} & {\footnotesize 50788} & {\footnotesize 50833} &  {\footnotesize 51116} & {\footnotesize } &  {\footnotesize } & {\footnotesize } & 
 {\footnotesize } & {\footnotesize  } \\
{\footnotesize 03371+4932} & {\footnotesize 51116} & {\footnotesize } & {\footnotesize } &  {\footnotesize } &  {\footnotesize } & {\footnotesize } & {\footnotesize } &  {\footnotesize } & {\footnotesize } &  {\footnotesize } & {\footnotesize } & 
 {\footnotesize } & {\footnotesize  } \\
{\footnotesize 03469+5833} & {\footnotesize 50711} & {\footnotesize 50735} & {\footnotesize 50742} &  {\footnotesize 50763} &  {\footnotesize 50833} & {\footnotesize 51116} & {\footnotesize } &  {\footnotesize } & {\footnotesize } &  {\footnotesize } & {\footnotesize } & 
 {\footnotesize } & {\footnotesize  } \\
{\footnotesize 03525+5711} & {\footnotesize 50466} & {\footnotesize 50672} & {\footnotesize 50694} &  {\footnotesize 50711} &  {\footnotesize 50735} & {\footnotesize 50742} & {\footnotesize 50763} &  {\footnotesize 50830} & {\footnotesize 50891} &  {\footnotesize 51116} & {\footnotesize } & 
 {\footnotesize } & {\footnotesize  } \\
{\footnotesize 03557+4404} & {\footnotesize 50712} & {\footnotesize 50833} & {\footnotesize 51116} &  {\footnotesize } &  {\footnotesize } & {\footnotesize } & {\footnotesize } &  {\footnotesize } & {\footnotesize } &  {\footnotesize } & {\footnotesize } & 
 {\footnotesize } & {\footnotesize  } \\
{\footnotesize 04085+5347} & {\footnotesize 50466} & {\footnotesize 50468} & {\footnotesize 50676} &  {\footnotesize 50694} &  {\footnotesize 50711} & {\footnotesize 50735} & {\footnotesize 50742} &  {\footnotesize 50763} & {\footnotesize 50788} &  {\footnotesize 50830} & {\footnotesize 50891} & 
 {\footnotesize 51116} & {\footnotesize  } \\
{\footnotesize 04209+4800} & {\footnotesize 50676} & {\footnotesize 50694} & {\footnotesize 50742} &  {\footnotesize 50763} &  {\footnotesize 50788} & {\footnotesize 50833} & {\footnotesize 51116} &  {\footnotesize } & {\footnotesize } &  {\footnotesize } & {\footnotesize } & 
 {\footnotesize } & {\footnotesize  } \\
{\footnotesize 04402+3426} & {\footnotesize 50694} & {\footnotesize 50711} & {\footnotesize 50735} &  {\footnotesize 50742} &  {\footnotesize 50763} & {\footnotesize 50830} & {\footnotesize 50891} &  {\footnotesize 51116} & {\footnotesize 51118} &  {\footnotesize } & {\footnotesize } & 
 {\footnotesize } & {\footnotesize  } \\
{\footnotesize 04470+3002} & {\footnotesize 50735} & {\footnotesize 50742} & {\footnotesize 50763} &  {\footnotesize 50833} &  {\footnotesize 50891} & {\footnotesize 51116} & {\footnotesize 51118} &  {\footnotesize } & {\footnotesize } &  {\footnotesize } & {\footnotesize } & 
 {\footnotesize } & {\footnotesize  } \\
{\footnotesize 05091+4639} & {\footnotesize 50694} & {\footnotesize 50735} & {\footnotesize 50742} &  {\footnotesize 50785} &  {\footnotesize 50830} & {\footnotesize 50879} & {\footnotesize 51118} &  {\footnotesize } & {\footnotesize } &  {\footnotesize } & {\footnotesize } & 
 {\footnotesize } & {\footnotesize  } \\
{\footnotesize 05146+2521} & {\footnotesize 50742} & {\footnotesize 50785} & {\footnotesize 50833} &  {\footnotesize 50879} &  {\footnotesize 51118} & {\footnotesize } & {\footnotesize } &  {\footnotesize } & {\footnotesize } &  {\footnotesize } & {\footnotesize } & 
 {\footnotesize } & {\footnotesize  } \\
{\footnotesize 05204+3227} & {\footnotesize 50785} & {\footnotesize 50830} & {\footnotesize 50891} &  {\footnotesize 51118} &  {\footnotesize } & {\footnotesize } & {\footnotesize } &  {\footnotesize } & {\footnotesize } &  {\footnotesize } & {\footnotesize } & 
 {\footnotesize } & {\footnotesize  } \\
{\footnotesize 05273+2019} & {\footnotesize 50785} & {\footnotesize 50833} & {\footnotesize 50879} &  {\footnotesize 51118} &  {\footnotesize } & {\footnotesize } & {\footnotesize } &  {\footnotesize } & {\footnotesize } &  {\footnotesize } & {\footnotesize } & 
 {\footnotesize } & {\footnotesize  } \\
{\footnotesize 05405+3240} & {\footnotesize 50830} & {\footnotesize 50891} & {\footnotesize 51118} &  {\footnotesize } &  {\footnotesize } & {\footnotesize } & {\footnotesize } &  {\footnotesize } & {\footnotesize } &  {\footnotesize } & {\footnotesize } & 
 {\footnotesize } & {\footnotesize  } \\
{\footnotesize 05423+2905} & {\footnotesize 50742} & {\footnotesize 50785} & {\footnotesize 50833} &  {\footnotesize 50879} &  {\footnotesize 51118} & {\footnotesize } & {\footnotesize } &  {\footnotesize } & {\footnotesize } &  {\footnotesize } & {\footnotesize } & 
 {\footnotesize } & {\footnotesize  } \\
{\footnotesize 05452+2001} & {\footnotesize 50830} & {\footnotesize 50891} & {\footnotesize 51118} &  {\footnotesize } &  {\footnotesize } & {\footnotesize } & {\footnotesize } &  {\footnotesize } & {\footnotesize } &  {\footnotesize } & {\footnotesize } & 
 {\footnotesize } & {\footnotesize  } \\
{\footnotesize 05484+3521} & {\footnotesize 50742} & {\footnotesize 50785} & {\footnotesize 50833} &  {\footnotesize 50879} &  {\footnotesize 51118} & {\footnotesize } & {\footnotesize } &  {\footnotesize } & {\footnotesize } &  {\footnotesize } & {\footnotesize } & 
 {\footnotesize } & {\footnotesize  } \\
{\footnotesize 05552+1720} & {\footnotesize 50785} & {\footnotesize 50830} & {\footnotesize 50891} &  {\footnotesize 51130} &  {\footnotesize } & {\footnotesize } & {\footnotesize } &  {\footnotesize } & {\footnotesize } &  {\footnotesize } & {\footnotesize } & 
 {\footnotesize } & {\footnotesize  } \\
{\footnotesize 06170+3523} & {\footnotesize 50735} & {\footnotesize 50785} & {\footnotesize 50830} &  {\footnotesize 50879} &  {\footnotesize 51130} & {\footnotesize } & {\footnotesize } &  {\footnotesize } & {\footnotesize } &  {\footnotesize } & {\footnotesize } & 
 {\footnotesize } & {\footnotesize  } \\
{\footnotesize 06447+0817} & {\footnotesize 50785} & {\footnotesize 50830} & {\footnotesize 50891} &  {\footnotesize 51130} &  {\footnotesize } & {\footnotesize } & {\footnotesize } &  {\footnotesize } & {\footnotesize } &  {\footnotesize } & {\footnotesize } & 
 {\footnotesize } & {\footnotesize  } \\
{\footnotesize 06448+1639} & {\footnotesize } & {\footnotesize } & {\footnotesize } &  {\footnotesize } &  {\footnotesize } & {\footnotesize } & {\footnotesize } &  {\footnotesize } & {\footnotesize } &  {\footnotesize } & {\footnotesize } & 
 {\footnotesize } & {\footnotesize  } \\
{\footnotesize 20523+5302} & {\footnotesize 50710} & {\footnotesize 50712} & {\footnotesize 50728} &  {\footnotesize 50742} &  {\footnotesize 50749} & {\footnotesize 50772} & {\footnotesize 50879} &  {\footnotesize 51096} & {\footnotesize } &  {\footnotesize } & {\footnotesize } & 
 {\footnotesize } & {\footnotesize  } \\
{\footnotesize 20532+5554} & {\footnotesize 50702} & {\footnotesize 50710} & {\footnotesize 50749} &  {\footnotesize 50772} &  {\footnotesize 51096} & {\footnotesize } & {\footnotesize } &  {\footnotesize } & {\footnotesize } &  {\footnotesize } & {\footnotesize } & 
 {\footnotesize } & {\footnotesize  } \\
{\footnotesize 21086+5238} & {\footnotesize 50662} & {\footnotesize 50676} & {\footnotesize 50702} &  {\footnotesize 50710} &  {\footnotesize 50712} & {\footnotesize 50728} & {\footnotesize 50739} &  {\footnotesize 50749} & {\footnotesize 50772} &  {\footnotesize 50788} & {\footnotesize 50879} & 
 {\footnotesize 51096} & {\footnotesize  } \\
{\footnotesize 21216+5536} & {\footnotesize 50662} & {\footnotesize 50676} & {\footnotesize 50702} &  {\footnotesize 50710} &  {\footnotesize 50728} & {\footnotesize 50739} & {\footnotesize 50749} &  {\footnotesize 50772} & {\footnotesize 50788} &  {\footnotesize 51096} & {\footnotesize } & 
 {\footnotesize } & {\footnotesize  } \\
{\footnotesize 21377+5042} & {\footnotesize 50749} & {\footnotesize 50772} & {\footnotesize 50788} &  {\footnotesize 5109} &  {\footnotesize } & {\footnotesize } & {\footnotesize } &  {\footnotesize } & {\footnotesize } &  {\footnotesize } & {\footnotesize } & 
 {\footnotesize } & {\footnotesize  } \\
{\footnotesize 21415+5025} & {\footnotesize 50662} & {\footnotesize 50676} & {\footnotesize 50702} &  {\footnotesize 50710} &  {\footnotesize 50728} & {\footnotesize 50739} & {\footnotesize 50749} &  {\footnotesize 50772} & {\footnotesize 50788} &  {\footnotesize 51112} & {\footnotesize } & 
 {\footnotesize } & {\footnotesize  } \\
{\footnotesize 21449+4950} & {\footnotesize } & {\footnotesize } & {\footnotesize } &  {\footnotesize } &  {\footnotesize } & {\footnotesize } & {\footnotesize } &  {\footnotesize } & {\footnotesize } &  {\footnotesize } & {\footnotesize } & 
 {\footnotesize } & {\footnotesize  } \\
{\footnotesize 21453+5959} & {\footnotesize } & {\footnotesize } & {\footnotesize } &  {\footnotesize } &  {\footnotesize } & {\footnotesize } & {\footnotesize } &  {\footnotesize } & {\footnotesize } &  {\footnotesize } & {\footnotesize } & 
 {\footnotesize } & {\footnotesize  } \\
{\footnotesize 21509+6234} & {\footnotesize 50110} & {\footnotesize 50468} & {\footnotesize 50662} &  {\footnotesize 50676} &  {\footnotesize 50702} & {\footnotesize 50711} & {\footnotesize 50728} &  {\footnotesize 50739} & {\footnotesize 50749} &  {\footnotesize 50785} & {\footnotesize 51112} & 
 {\footnotesize } & {\footnotesize  } \\
{\footnotesize 21563+5630} & {\footnotesize 50676} & {\footnotesize 50702} & {\footnotesize 50711} &  {\footnotesize 50728} &  {\footnotesize 50739} & {\footnotesize 50749} & {\footnotesize 50785} &  {\footnotesize 51112} & {\footnotesize } &  {\footnotesize } & {\footnotesize } & 
 {\footnotesize } & {\footnotesize  } \\
{\footnotesize 22241+6005} & {\footnotesize 50785} & {\footnotesize 50833} & {\footnotesize 50879} &  {\footnotesize 51118} &  {\footnotesize } & {\footnotesize } & {\footnotesize } &  {\footnotesize } & {\footnotesize } &  {\footnotesize } & {\footnotesize } & 
 {\footnotesize } & {\footnotesize  } \\
{\footnotesize 22394+5623} & {\footnotesize 50739} & {\footnotesize 50760} & {\footnotesize 50788} &  {\footnotesize 50833} &  {\footnotesize 51112} & {\footnotesize 51116} & {\footnotesize } &  {\footnotesize } & {\footnotesize } &  {\footnotesize } & {\footnotesize } & 
 {\footnotesize } & {\footnotesize  } \\
{\footnotesize 22394+6930} & {\footnotesize 50711} & {\footnotesize 50728} & {\footnotesize 50739} &  {\footnotesize 50749} &  {\footnotesize 50785} & {\footnotesize 50833} & {\footnotesize 51112} &  {\footnotesize } & {\footnotesize } &  {\footnotesize } & {\footnotesize } & 
 {\footnotesize } & {\footnotesize  } \\
{\footnotesize 22466+6942} & {\footnotesize 50702} & {\footnotesize 50711} & {\footnotesize 50739} &  {\footnotesize 50760} &  {\footnotesize 50788} & {\footnotesize 50833} & {\footnotesize 51116} &  {\footnotesize } & {\footnotesize } &  {\footnotesize } & {\footnotesize } & 
 {\footnotesize } & {\footnotesize  } \\
{\footnotesize 23491+6243} & {\footnotesize } & {\footnotesize } & {\footnotesize } &  {\footnotesize } &  {\footnotesize } & {\footnotesize } & {\footnotesize } &  {\footnotesize } & {\footnotesize } &  {\footnotesize } & {\footnotesize } & 
 {\footnotesize } & {\footnotesize  } \\

\\[4pt]
\hline
\end{tabular*}
 
\vspace{6pt}

\noindent

\end{table*}

\begin{table*}
\renewcommand{\baselinestretch}{1.0}
\small
\begin{center}
Table~2.\hspace{4pt}Results of photometric obsevation and PDM analysis.\\
\end{center}
\vspace{6pt}
\begin{tabular*}{\textwidth}{@{\hspace{\tabcolsep}
\extracolsep{\fill}}p{6pc}cccccccc}
\hline\hline\\[-6pt]
IRAS NAME & $<I>$ & $\Delta I/2$ & No.of Obs. & Base & Period & $\Delta P.$ & Epoch \\
[4pt]\hline\\[-6pt]
 & (mag) & (mag) & & (day) & (day) & (day) & (JD-2400000) \\
00336+6744 & \phantom{0}9.1 & 1.7 & 23 & 1412 & 212 & \phantom{0}16 & 49679 \\
00459+6749 & 11.0 & 0.9 & 21 & 1412 & \phantom{0}180$^{*}$ & \phantom{0}11 & 49575 \\
00534+6031 & 10.5 & 1.5 & 12 & 1463 & \phantom{0}410$^{*}$ & \phantom{0}57 & 49816 \\
00589+5743 & \phantom{0}8.6 & 0.4 & 22 & 1466 & [188] & \phantom{0}[12] & [49648] \\
02117+5559 & 15.5 & 2.4 & 21 & 1466 & \phantom{0}509$^{*}$ & \phantom{0}88 & 49891 \\
02272+6327 & 15.3 & 1.0 & 23 & 1466 & 572 & 112 & 49593 \\
02433+6345 & 10.5 & 1.6 & 24 & 1508 & 460 & \phantom{0}70 & 49763 \\
02470+5536 & 11.1 & 1.2 & 10 & 1467 & \phantom{0}480$^{*}$ & \phantom{0}79 & 49620 \\
03022+5409 & 12.8 & 1.8 & 25 & 1463 & 582 & 116 & 49944 \\
03096+5936 & 15.9 & 2.0 & 24 & 1509 & \phantom{0}781$^{*}$ & 202 & 49922 \\
03192+5642 & 13.7 & 1.0 & 24 & 1509 & \phantom{0}430$^{*}$ & \phantom{0}61 & 50004 \\
03238+6034 & 16.2 & 1.3 & 21 & 1489 & \phantom{0}477$^{*}$ & \phantom{0}76 & 50047 \\
03371+4932 & 17.6 & 1.5 & 14 & 1489 & [127] & \phantom{00}[5] & [49630] \\
03469+5833 & 11.7 & 1.8 & 19 & 1456 & 617 & 131 & 49823 \\
03525+5711 & 12.3 & 2.2 & 23 & 1489 & \phantom{0}588$^{*}$ & 116 & 49885 \\
03557+4404 & 17.6 & 1.4 & 16 & 1489 & [482] & \phantom{0}[78] & [49821] \\
04085+5347 & \phantom{0}9.4 & 2.1 & 25 & 1489 & 486 & \phantom{0}79 & 49638 \\
04209+4800 & 12.8 & 1.8 & 20 & 1456 & \phantom{0}788$^{*}$ & 213 & 50292 \\
04402+3426 & 11.6 & 1.7 & 22 & 1457 & 386 & \phantom{0}51 & 49952 \\
04470+3002 & 11.9 & 1.6 & 20 & 1458 & 404 & \phantom{0}56 & 49946 \\
05091+4639 & 10.1 & 1.6 & 20 & 1458 & \phantom{0}433$^{*}$ & \phantom{0}64 & 50050 \\
05146+2521 & 11.5 & 1.8 & 18 & 1458 & \phantom{0}536$^{*}$ & \phantom{0}99 & 49682 \\
05204+3227 & 15.2 & 2.0 & 17 & 1457 & 575 & 114 & 49935 \\
05273+2019 & 13.1 & 1.9 & 17 & 1431 & 446 & \phantom{0}69 & 49642 \\
05405+3240 & 17.0 & 1.1 & 16 & 1457 & \phantom{0}683$^{*}$ & 160 & 49611 \\
05423+2905 & 11.2 & 1.5 & 18 & 1458 & 449 & \phantom{0}69 & 49955 \\
05452+2001 & 15.1 & 1.0 & 16 & 1431 & 660 & 152 & 49835 \\
05484+3521 & 14.5 & 1.0 & 18 & 1457 & 535 & \phantom{0}98 & 49781 \\
05552+1720 & 13.6 & 2.1 & 17 & 1443 & \phantom{0}218$^{*}$ & \phantom{0}17 & 49777 \\
06170+3523 & 13.4 & 2.1 & 18 & 1470 & 428 & \phantom{0}62 & 49708 \\
06447+0817 & 15.6 & 1.7 & 17 & 1444 & \phantom{0}476$^{*}$ & \phantom{0}78 & 49890 \\
06448+1639 & 11.8 & 1.2 & 13 & 1444 & \phantom{0}498$^{*}$ & \phantom{0}86 & 49815 \\
20523+5302 & 14.6 & 1.9 & 21 & 1531 & \phantom{0}500$^{*}$ & \phantom{0}82 & 49803 \\
20532+5554 & 17.4 & 1.3 & 18 & 1534 & [591] & [114] & [50190] \\
21086+5238 & 10.6 & 1.3 & 25 & 1534 & \phantom{0}400$^{*}$ & \phantom{0}52 & 49763 \\
21216+5536 & 10.4 & 1.3 & 23 & 1531 & 421 & \phantom{0}58 & 49856 \\
21377+5042 & 15.8 & 1.3 & 17 & 1534 & \phantom{0}500$^{*}$ & \phantom{0}82 & 50050 \\
21415+5025 & \phantom{0}9.9 & 2.2 & 23 & 1550 & \phantom{0}416$^{*}$ & \phantom{0}56 & 49635 \\
21449+4950 & 16.4 & 0.8 & 12 & 1550 & \phantom{0}646$^{*}$ & 135 & 50118 \\
21453+5959 & 15.2 & 1.5 & 13 & 1220 & \phantom{0}537$^{*}$ & 118 & 49650 \\
21509+6234 & 10.4 & 2.1 & 24 & 1547 & 512 & \phantom{0}85 & 49964 \\
21563+5630 & \phantom{0}9.0 & 1.7 & 21 & 1547 & 523 & \phantom{0}88 & 49900 \\
22241+6005 & 14.5 & 1.9 & 17 & 1431 & \phantom{0}698$^{*}$ & 170 & 50095 \\
22394+5623 & 10.3 & 1.7 & 19 & 1455 & 458 & \phantom{0}72 & 49589 \\
22394+6930 & 11.1 & 1.2 & 20 & 1451 & \phantom{0}183$^{*}$ & \phantom{0}12 & 49635 \\
22466+6942 & \phantom{0}9.2 & 1.7 & 20 & 1551 & \phantom{0}408$^{*}$ & \phantom{0}54 & 49617 \\
23491+6243 & 17.0 & 3.1 & 13 & 1554 & \phantom{0}575$^{*}$ & 106 & 50046 \\

\\[4pt]
\hline
\end{tabular*}

\vspace{6pt}

\noindent
$^*$ {\it Uncertain periods.\\}
[ ] {\it The numbers in brackets are the data of period undetermined objects. Therefore, reliability of 
these data is quite low.} 
\end{table*}

\begin{table*}
\renewcommand{\baselinestretch}{1.0}
\small
\begin{center}
Table~3.\hspace{4pt}Results of luminosity calculations, IRAS flux and radio data.\\
\end{center}
\vspace{6pt}
\begin{tabular*}{\textwidth}{@{\hspace{\tabcolsep}
\extracolsep{\fill}}p{6pc}cccccccc}
\hline\hline\\[-6pt]
IRAS NAME & $<I>_{abs}$ & $<M_{bol}>_{abs}$ & $F_{12}$ & $C_{12}$ & $C_{23}$ & $V_{lsr}$(SiO) & CSE \\
[4pt]\hline\\[-6pt]
 & (mag) & (mag) & ($Jy$) & & & ($km/s$) & \\
00336+6744$^*$ & -4.4 & -4.2 & \phantom{00}6.373 & -0.20 & -0.77 & \phantom{0}-91.0 & O \\
00459+6749 & -4.3 & -4.0 & \phantom{00}8.069 & -0.28 & -0.77 & \phantom{0}-89.0 & O \\
00534+6031 & -4.7 & -5.0 & \phantom{00}4.691 & -0.20 & -0.46 & -101.7 & O \\
00589+5743 & -4.3 & -4.0 & \phantom{00}4.284 & -0.23 & -0.73 &  & O \\
02117+5559 & -4.8 & -5.6 & \phantom{00}7.399 & -0.04 & -0.86 &  & O \\
02272+6327$^*$ & -4.9 & -6.0 & \phantom{00}8.261 & -0.29 & -0.62 &  & unknown \\
02433+6345$^*$ & -4.8 & -5.3 & \phantom{0}20.080 & -0.08 & -0.85 & \phantom{0}-56.2 & O \\
02470+5536 & -4.8 & -5.4 & \phantom{0}29.280 & -0.20 & -0.92 & \phantom{0}-37.0 & O \\
03022+5409$^*$ & -4.9 & -6.1 & \phantom{0}11.160 & -0.17 & -0.87 & \phantom{000}-28.7$^{**}$ & O \\
03096+5936 &  & -5.1 & \phantom{0}16.330 & -0.21 & -0.54 &  & $C$ \\
03192+5642 &  & -4.6 & \phantom{0}16.880 & -0.25 & -0.40 &  & $C$ \\
03238+6034 &  & -4.7 & \phantom{0}70.430 & -0.27 & -0.67 &  & $C$ \\
03371+4932 & -4.1 & -3.5 & \phantom{00}5.420 & -0.27 & -0.62 &  & unknown \\
03469+5833$^*$ & -4.9 & -6.3 & \phantom{00}7.343 & -0.26 & -0.72 & \phantom{0}-74.3 & O \\
03525+5711 & -4.9 & -6.1 & \phantom{00}7.132 & -0.21 & -0.70 &  & O \\
03557+4404 &  & -4.7 & \phantom{0}36.300 & -0.09 & -0.56 &  & $C$ \\
04085+5347$^*$ & -4.8 & -5.5 & \phantom{0}29.530 & -0.26 & -0.78 & \phantom{0}-15.3 & O \\
04209+4800 & -5.1 & -7.1 & \phantom{0}28.870 & -0.11 & -0.77 & \phantom{0}-16.3 & O \\
04402+3426$^*$ & -4.7 & -4.9 & \phantom{00}3.394 & -0.23 & -0.65 &  & O \\
04470+3002$^*$ & -4.7 & -5.0 & \phantom{00}5.950 & -0.12 & -0.73 &  & O \\
05091+4639 & -4.8 & -5.1 & \phantom{0}17.210 & -0.17 & -0.77 & \phantom{00}12.7 & O \\
05146+2521 & -4.9 & -5.8 & \phantom{0}19.960 & -0.22 & -0.87 & \phantom{00}-1.8 & O \\
05204+3227$^*$ & -4.9 & -6.0 & \phantom{0}23.010 & -0.17 & -0.89 &  & O \\
05273+2019$^*$ &  & -4.7 & \phantom{00}5.242 & -0.17 & -0.60 &  & $C$ \\
05405+3240 &  & -5.0 & 196.000 & -0.19 & -0.57 &  & $C$ \\
05423+2905$^*$ & -4.8 & -5.1 & \phantom{0}49.810 & -0.10 & -0.57 & \phantom{00}27.2 & O \\
05452+2001$^*$ &  & -5.0 & \phantom{0}15.740 & -0.21 & -0.69 &  & $C$ \\
05484+3521$^*$ &  & -4.8 & \phantom{00}8.001 & -0.29 & -0.55 &  & $C$ \\
05552+1720 & -4.4 & -4.2 & \phantom{00}8.673 & -0.10 & -0.72 & \phantom{00}43.6 & O \\
06170+3523$^*$ & -4.7 & -5.1 & \phantom{0}22.840 & -0.20 & -0.91 & \phantom{00}23.6 & O \\
06447+0817 & -4.8 & -5.4 & \phantom{0}21.250 & -0.28 & -0.65 &  & unknown \\
06448+1639 &  & -4.8 & \phantom{00}5.756 & -0.30 & -0.58 &  & $C$ \\
20523+5302 & -4.8 & -5.5 & \phantom{0}11.070 & -0.09 & -0.70 & \phantom{0}-66.3 & O \\
20532+5554 &  & -4.9 & \phantom{0}58.530 & -0.12 & -0.63 &  & $C$ \\
21086+5238 & -4.7 & -5.0 & \phantom{0}54.320 & -0.28 & -0.69 & \phantom{000}-13.4$^{**}$ & O \\
21216+5536$^*$ & -4.7 & -5.0 & \phantom{0}11.410 & -0.19 & -0.69 & \phantom{00}-9.8 & O \\
21377+5042 &  & -4.8 & \phantom{0}94.490 & -0.20 & -0.95 &  & $C$ \\
21415+5025 & -4.7 & -5.0 & \phantom{00}8.293 & -0.21 & -0.63 &  & O \\
21449+4950 &  & -5.0 & \phantom{0}67.070 & -0.18 & -0.71 &  & $C$ \\
21453+5959 & -4.9 & -5.8 & \phantom{0}24.380 & -0.11 & -0.69 & \phantom{0}-43.7 & O \\
21509+6234$^*$ & -4.8 & -5.6 & \phantom{0}23.990 & -0.22 & -0.72 & \phantom{0}-67.9 & O \\
21563+5630$^*$ & -4.9 & -5.7 & \phantom{0}84.300 & -0.21 & -0.66 & \phantom{000}-10.3$^{**}$ & O \\
22241+6005 &  & -5.0 & 181.500 & -0.23 & -0.62 &  & $C$ \\
22394+5623$^*$ & -4.8 & -5.3 & \phantom{00}6.833 & -0.26 & -0.67 & \phantom{0}-25.7 & O \\
22394+6930 & -4.3 & -4.0 & \phantom{00}5.606 & -0.26 & -0.71 & \phantom{0}-41.1 & O \\
22466+6942 & -4.7 & -5.0 & \phantom{0}16.180 & -0.19 & -0.79 & \phantom{0}-48.1 & O \\
23491+6243 &  & -4.9 & \phantom{0}40.610 & -0.27 & -0.48 &  & $C$ \\
\\[4pt]
\hline
\end{tabular*}

\vspace{6pt}

\noindent
$^*$ {\it These objects satisfied the period determination criteria. (see text)}\\
$^{**}$ {\it Newly detected SiO maser sources (Nakashima et al. 1999. to be published)} 

\end{table*}

\begin{table*}
\renewcommand{\baselinestretch}{1.0}
\small
\begin{center}
Table~4.\hspace{4pt}Results of distance calculation\\
\end{center}
\vspace{6pt}
\begin{tabular*}{\textwidth}{@{\hspace{\tabcolsep}
\extracolsep{\fill}}p{6pc}ccccccc}
\hline\hline\\[-6pt]
 & \multicolumn{4}{c}{METHOD 1} & \multicolumn{2}{c}{METHOD 2} \\ \cline{2-5}\cline{6-7}
IRAS NAME & $D_{MODEL1}$ & $A_{I1}$ & $D_{MODEL2}$ & $A_{I2}$ & $D_{bol}$ &
$BC_{12}$ \\
[4pt]\hline\\[-6pt]
 & (kpc) & (mag) & (kpc) & (mag) & (kpc) & \\
00336+6744 & \phantom{0}2.6 & 1.4 & \phantom{0}2.5 & 1.6 & 2.3 & 14.3 \\
00459+6749 & \phantom{0}4.1 & 2.2 & \phantom{0}4.1 & 2.2 & 1.4 & 24.9 \\
00534+6031 & \phantom{0}4.0 & 2.2 & \phantom{0}4.0 & 2.2 & 3.8 & 14.3 \\
00589+5743 & \phantom{0}2.2 & 1.2 & \phantom{0}2.1 & 1.4 & 2.3 & 17.6 \\
02117+5559 & \phantom{0}8.5 & 5.7 & \phantom{0}26.1$^*$ & 3.2 & 6.5 & \phantom{0}5.5 \\
02272+6327 & \phantom{0}8.3 & 5.6 & \phantom{0}24.8$^*$ & 3.2 & 3.4 & 26.8 \\
02433+6345 & \phantom{0}3.7 & 2.5 & \phantom{0}4.3 & 2.1 & 3.1 & \phantom{0}6.8 \\
02470+5536 & \phantom{0}4.2 & 2.8 & \phantom{0}5.2 & 2.3 & 1.9 & 14.3 \\
03022+5409 & \phantom{0}5.6 & 3.9 & \phantom{0}9.5 & 2.8 & 4.6 & 11.7 \\
03096+5936 &  &  &  &  & 2.1 & 15.3 \\
03192+5642 &  &  &  &  & 1.4 & 20.2 \\
03238+6034 &  &  &  &  & 0.7 & 23.2 \\
03371+4932 & \phantom{0}9.7 & 6.8 & \phantom{0}51.4$^*$ & 2.9 & 1.4 & 23.2 \\
03469+5833 & \phantom{0}4.7 & 3.3 & \phantom{0}6.6 & 2.5 & 4.6 & 21.7 \\
03525+5711 & \phantom{0}5.2 & 3.6 & \phantom{0}8.2 & 2.6 & 5.0 & 15.3 \\
03557+4404 &  &  &  &  & 1.7 & \phantom{0}7.2 \\
04085+5347 & \phantom{0}2.8 & 2.0 & \phantom{0}3.2 & 1.7 & 1.6 & 21.7 \\
04209+4800 & \phantom{0}6.2 & 3.9 & \phantom{0}11.1$^*$ & 2.7 & 5.4 & \phantom{0}8.1 \\
04402+3426 & \phantom{0}5.3 & 2.7 & \phantom{0}6.4 & 2.3 & 3.9 & 17.6 \\
04470+3002 & \phantom{0}6.2 & 2.7 & \phantom{0}7.1 & 2.3 & 4.4 & \phantom{0}8.6 \\
05091+4639 & \phantom{0}3.9 & 2.0 & \phantom{0}4.0 & 1.9 & 2.3 & 11.7 \\
05146+2521 & \phantom{0}5.9 & 2.5 & \phantom{0}6.7 & 2.3 & 2.5 & 16.4 \\
05204+3227 & \phantom{0}11.3$^*$ & 4.8 & \phantom{0}31.0$^*$ & 2.6 & 3.0 & 11.7 \\
05273+2019 &  &  &  &  & 3.5 & 11.7 \\
05405+3240 &  &  &  &  & 0.6 & 13.4 \\
05423+2905 & \phantom{0}5.4 & 2.3 & \phantom{0}5.8 & 2.2 & 1.7 & \phantom{0}7.6 \\
05452+2001 &  &  &  &  & 2.0 & 15.3 \\
05484+3521 &  &  &  &  & 2.0 & 26.8 \\
05552+1720 & \phantom{0}9.6 & 3.1 & \phantom{0}12.1$^*$ & 2.6 & 2.7 & \phantom{0}7.6 \\
06170+3523 & \phantom{0}8.2 & 3.5 & \phantom{0}12.6$^*$ & 2.6 & 1.8 & 14.3 \\
06447+0817 & \phantom{0}14.1$^*$ & 4.7 & \phantom{0}33.1$^*$ & 2.8 & 1.6 & 24.9 \\
06448+1639 &  &  &  &  & 2.2 & 28.7 \\
20523+5302 & \phantom{0}7.9 & 4.9 & \phantom{0}8.7 & 4.7 & 4.5 & \phantom{0}7.2 \\
20532+5554 &  &  &  &  & 1.3 & \phantom{0}8.6 \\
21086+5238 & \phantom{0}3.8 & 2.4 & \phantom{0}3.6 & 2.5 & 0.9 & 24.9 \\
21216+5536 & \phantom{0}3.7 & 2.3 & \phantom{0}3.5 & 2.4 & 2.6 & 13.4 \\
21377+5042 &  &  &  &  & 0.8 & 14.3 \\
21415+5025 & \phantom{0}3.3 & 2.0 & \phantom{0}3.1 & 2.2 & 2.8 & 15.3 \\
21449+4950 &  &  &  &  & 1.1 & 12.5 \\
21453+5959 & \phantom{0}8.3 & 5.5 & \phantom{0}12.8$^*$ & 4.6 & 3.3 & \phantom{0}8.1 \\
21509+6234 & \phantom{0}3.6 & 2.4 & \phantom{0}3.7 & 2.4 & 2.1 & 16.4 \\
21563+5630 & \phantom{0}2.7 & 1.8 & \phantom{0}2.6 & 1.8 & 1.2 & 15.3 \\
22241+6005 &  &  &  &  & 0.6 & 17.6 \\
22394+5623 & \phantom{0}3.9 & 2.2 & \phantom{0}3.7 & 2.2 & 3.0 & 21.7 \\
22394+6930 & \phantom{0}3.8 & 2.5 & \phantom{0}3.9 & 2.4 & 1.8 & 21.7 \\
22466+6942 & \phantom{0}2.9 & 1.6 & \phantom{0}2.7 & 1.8 & 2.1 & 13.4 \\
23491+6243 &  &  &  &  & 1.0 & 23.2 \\
\\[4pt]
\hline
\end{tabular*}

\vspace{6pt}

\noindent
$^*$ {\it PDM solutions further than 10 kpc.(see text)}

\end{table*}

\section*{References}
\small

\re
Alves D., Alcock C., Marshall S., Minniti D., Allsman R., Axelrod T.,
 Freeman K., Peterson B.et al. \ 1998, IAUS 190,47

\re
Arimoto J., Sadakane K.\ 1996, Memoirs of Osaka Kyoiku Univ.(\,{\footnotesize III}),45(1),131

\re
Bedding T.R., Zijlstra A.A.\ 1998, ApJL 506,47

\re
Beichman C.A., Neugebauer G., Habing H.J., Clegg P.E., Chester T.J.\ 1985,
IRAS Catalogues \& Atlases Explanatory Supplement,US Government Printing
Office,Washinton DC

\re
Bessell M.S.\ 1990, PASP 102,1181

\re
Binney J., Dehnen W.\ 1997, MNRAS 287,L5

\re
Brand J., Blitz L.\ 1993, A\&A 275,67

\re
Burton W.B., Gordon M.A.\ 1978, A\&A 63,7

\re
Caldwell J.A.R., Coulson I.M.\ 1985, MNRAS 212,879

\re
Chen B., Vergely J.L.Valette B., Carraro G.\ 1998, A\&A 336,137

\re
Clemens D.P.\ 1985, ApJ 295,422

\re
Deguchi S., Matsumoto S., Wood P.R.\ 1998, PASJ 50,597

\re
Feast M.W.\ 1996, MNRAS 278,11

\re
Feast M.W., Glass I.S., Whitelock P.A., Catchpole R.M.\ 1989, MNRAS 241,375

\re
Fernie J.D.\ 1989, PASP 101,225

\re
Fich M., Blitz L., Stark A.A.\ 1989, ApJ 342,272

\re
Frink S., Fuchs B., Roeser S., Wielen R.\ 1996, A\&A 314,430

\re
Glass I.S., Whitelock P.A., Catchpole R.M., Feast M.W.\ 1995, MNRAS 273,383

\re
He L., Whittet D.C.B., Kilkenny D., Jones J.H.S.\ 1995, ApJS 101,335

\re
Hron J.\ 1987, A\&A 176,34

\re
Hughes S.M.G., Wood P.R.\ 1990, AJ 99,784

\re
Izumiura H.\ 1990, Ph.D.thesis,Univ.\ of Tokyo

\re
Izumiura H., Deguchi S., Hashimoto O., Nakada Y., Onaka T., Ono T., Ukita N., Yamamura I.\ 1994,  ApJ 437,419

\re
Jewell P.R., Snyder L.E., Walmsley C.M., Wilson T.L., Gensheimer P.D.\ 1991, A\&A 242,211

\re
Jiang B.W.\ 1998, PASP 110,214

\re
Jiang B.W., Deguchi S., Nakada Y.\ 1996, AJ 111,231

\re
Jiang B.W., Deguchi S., Hu J.Y., Yamashita T., Nishihara E., Matsumoto S., Nakada Y.\ 1997, AJ 113,1315

\re
Jiang B.W., Deguchi S., Yamamura I., Nakada Y., Cho S.H., Yamagata T.\ 1996, ApJS 106,463

\re
Kerr F.J., Lynden-Bell D.\ 1986, MNRAS 221,1023

\re
Landolt A.U.\ 1992, AJ 104,340

\re
Lee S.G., Kim E., Lee H.M.\ 1994, J.\ Korean Astro.\ Soc.\ 27,133

\re
Loup C., Forveille T., Omont A., Paul J.F.\ 1993, A\&AS 99, 291

\re
Merrifield M.R.\ 1992, AJ 103,1552

\re
Metzger M.R., Schechter P.L.\ 1994, ApJ 420,177

\re
Nakashima J., Deguchi S., Izumiura H.\ 1999, to be published

\re
Nemec A.F.L., Nemec J,M,\ 1985, AJ 90,2317

\re
Olnon F.M., Raimond E., and IRAS Science Team \ 1986, A\&AS 65,607

\re
Pont F., Mayor M., Burki G.\ 1994, A\&A 285,415

\re
Reid N., Glass I.S., Catchpole R.M.\ 1988, MNRAS 232,53

\re
Schneider S.E., Terzian Y.\ 1983, ApJL 274,L61

\re
Sofue Y.\ astro-ph/9808011

\re
Stellingwerf R.F.\ 1978, ApJ 224,953

\re
te Lintel-Hekkert P., Versteege-Hensel H.A., Habing H.J., Wiertz M.\ 1989, A\&AS 78,399

\re
Unavane M., Gilmore G., Epchtein N., Simon G., Tiphene D., de Batz B.\ 1998, MNRAS 295,119

\re
van der Veen W.E.C.J., Breukers R.J.L.H.\ 1989, A\&A 213,133

\re
van Leeuwen F., Feast M.W., Whitelock P.A., Yudin B.\ 1997, MNRAS 287,955

\re
Walker G.\ 1987, Astronomical Observations, Cambridge Univ.\ Press,p.\ 47

\re
Whitelock P.A., Feast M.W., Catchpole R.M.\ 1991, MNRAS 248,276

\re
Whitelock P.A., Menzies J., Feast M.W., Marang F., Carter B., Roberts G., Catchpole R.M., Chapman J.\ 1994, MNRAS 267,711

\re
Wood P.R., Sebo K.M.\ 1996, MNRAS 282,958

\re
Zombeck M.V.\ 1982, Hand Book of Space Astronomy \& Astrophysics, Cambridge Univ.\ Press,p.\ 102 

\label{last}
\end{document}